\documentclass[fleqn,11pt]{wlscirep}
\usepackage[utf8]{inputenc}
\usepackage[T1]{fontenc}
\usepackage{float}
\usepackage{hyperref}
\title{Simplicial complex entropy for time series analysis}

\author[1*]{L. Guzmán-Vargas}
\author[1]{A. Zabaleta-Ortega}
\author[2]{A. Guzmán-Sáenz}

\affil[1,*]{Unidad Profesional Interdisciplinaria en Ingeniería y Tecnologías Avanzadas, Instituto Politécnico Nacional, 07340 Ciudad de México, México, azabaletao1900@alumno.ipn.mx, lguzmanv@ipn.mx}
\affil[2]{Topological Data Analysis in Genomics, Thomas J. Watson Research Center, Yorktown Heights, NY USA, Aldo.Guzman.Saenz@ibm.com}

\begin{abstract}
The complex behavior of many systems in nature requires the application of robust methodologies capable of identifying changes in their dynamics. In the case of time series (which are sensed values of a system during a time interval), several methods have been proposed to evaluate their irregularity. However, for some types of dynamics  such as stochastic and chaotic, new approaches are required that can provide a better characterization of them.  In this paper we present the  simplicial complex approximate entropy (SCAE), which is based on the conditional probability of the occurrence of elements of a simplicial complex. Our results show that this entropy measure provides a wide range of values with details not easily identifiable with standard methods. In particular, we show that our method is able to quantify the irregularity in simulated random sequences and those from low-dimensional chaotic dynamics. Furthermore, it is possible to consistently differentiate cardiac interbeat sequences from healthy subjects and from patients with heart failure, as well as to identify changes between dynamical states of coupled chaotic maps.  Our results highlight the importance of the structures revealed by the simplicial complexes, which holds promise for applications of this approach in various contexts.
\end{abstract}
\begin{document}

\flushbottom
\maketitle
%
%
\thispagestyle{empty}

\section{Introduction}\label{Introduction}
In recent years, various methodologies have been proposed to estimate the entropy of a system \cite{Kolmogorov1959, Renyi1961, Sinai1959,  Karmeshu2003, Beck2009, DehmerAndMowshowitz2011, ChenAndWang2016, Delgado_BonalAndMarshak2019, Namdari2019}. Some of these approaches have been especially useful when applied to real-world time series, including research areas such as medicine \cite{Pincus2006, bassingthwaighte1996fractal, CostaPRE2005, Garcia-MartinezEtAl2016, Keshmiri2020}, finance \cite{Pincus2006b, ZhouEtAl2013, KukretiEtAl2020}, mechanical systems \cite{HuoEtAl2020}, among others. 
In the context of dynamical systems, methods like the approximate entropy (ApEn) \cite{Pincus1991} and its derivatives have been introduced to measure the irregularity in a time series \cite{PincusAndHuang1992, Pincus1995,RichmanAndMoorman2000, XieEtAl2010, LiangEtAl2015}. 
ApEn measures the irregularity in a sequence by quantifying the repetition of patterns of a certain length in relation to when the length is increased by one value; a lower value of ApEn indicates a more regular behavior whereas a higher value means more irregularity. For a detailed description of ApEn and its modified versions, see \cite{Delgado_BonalAndMarshak2019, JaminEtAl2020, Humeau2015}. 
On the other hand, the use of tools from the so-called topological data analysis \cite{EdelsbrunnerEtAl2002, CarlssonZomorodian2007, EpsteinEtAl2011, Zomorodian2012, Munch2017, Wasserman2018, Motta2018, XuEtAl2019, AtienzaEtAl2020, Carlsson2020, ChazalAndMichel2021}
to characterize data whose complexity extends into higher order interactions has been gaining relevance \cite{Carlsson2020}. One of the main tools of TDA is the so-called persistent homology (PH), which provides information about the structure of the data \cite{ZomorodianAndCarlsson2004, CarlssonEtAl2005, EdelsbrunnerAndHarer2008, Carlsson2009, EdelsbrunnerAndMorozov2013, BerwaldEtAl2014, FasyEtAl2014, Lacasa2015}. In PH, we construct a point cloud and convert it into a collection of simplices to form simplicial complexes. Then one can calculate the homology of each of these complexes. These methodologies have been successfully applied to a number of data from very different contexts. 
In particular, applications of PH have been reported in neuroscience \cite{LeeEtAl2011, CaputiEtAl2021, NielsonEtAl2015, Curto2017, SaggarEtAl2018, SizemoreEtAl2018, SkafAndLaubenbacher2022}, image processing \cite{ChristianEtAl2022, Edelsbrunner2013}, DNA \cite{CamaraEtAl2016, AmezquitaEtAl2020, NielsonEtAl2017, HumphreysEtAl2019, ShoemarkEtAl2021}, among other fields \cite{MittalAndGupta2017, CarlssonAndVejdemo-Johansson2021}.
However, homology has not been widely used in the context of dynamical systems and in entropy-based analysis of the short and noisy data sets encountered in real studies. Of particular interest is the estimation of entropy in irregular series that often appear in real systems and whose estimation is crucial to characterize them. Also, these records are obtained in a bivariate or multivariate way, so it is useful to have these methodologies applicable in these cases. 
Despite the success of methodologies such as ApEn (and its variants), the range of applicability of these measures remains limited, because different dynamics may eventually lead to similar entropy values and not have a clear separation. 
In order to have a more robust measure and fill the gap, in this work we introduce the simplicial complex approximate entropy (SCAE) which is based on the ApEn proposed by Pincus \cite{Pincus1991}, and its improved versions \cite{PincusAndKeefe1992, PincusAndHuang1992, Pincus1995, Pincus2006, Pincus2006b, RichmanAndMoorman2000, ChenEtAl2006}.  For a cloud of points (patterns) for which a simplicial complex is defined, the SCAE is based on the conditional probability that two elements of the simplicial complex (of the same dimension $k$) that are "close" remain "close" with a third element, forming a simplex of a higher dimension ($k+1$). Unlike ApEn, the SCAE definition does not rely on increasing the length of the patterns to search for matches, but is based on the matches that exist between patterns of the same length when the simplex dimension is increased by one.
We show that this entropy measure provides a new way to identify details in signal complexity that are not easily quantified by standard methods, and extend our method to evaluate the level of synchrony between two signals. 
Specifically, we demonstrate that our method is capable of quantify irregularity in random sequences and low-dimensional chaotic dynamics, and that when applied to real-world series, it can clearly differentiate cardiac interbeat dynamics from healthy individuals and those with certain cardiac pathology. In addition, when applied to the bivariate case, it can identify changes between dynamic states of coupled chaotic maps.




\begin{figure}[H]
     \centering
\includegraphics[width=0.8\linewidth]{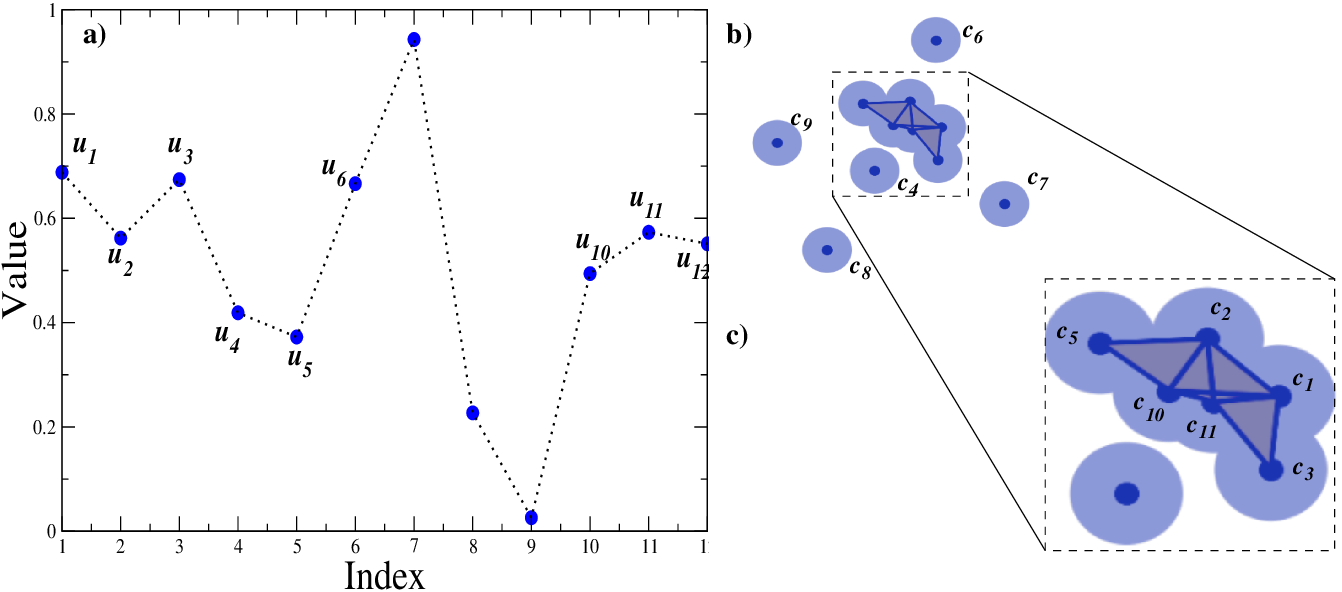}
    \caption{Representative time series from random numbers to illustrate the procedure for calculating the simplicial complex approximate entropy (SCAE).  Given the time series $u_1,u_2,..u_{12}$, the $2$-dimensional vectors $c_i=[u_i,u_{i+1}]$, with $1\leq i \leq11$, are constructed. b) For a given radius $\epsilon$, the Vietoris-Rips complex is created. Here we show the case $\epsilon=0.2$, denoted by the border of the shaded area in light blue. For the 11 points shown, there are $10$ edges and  $6$ triangles. Specifically, we have the following 1-simplexes: $\{c_1 c_3\},\{c_1 c_2\},\{c_1 c_{11}\}, \{c_1 c_{10}\},\{c_2 c_5\},\{c_2 c_{10}\}, \{c_2 c_{11}\},\{c_5 c_{10}\},\{c_{10} c_{11}\},\{c_{11} c_{3}\}$; and the following 2-simplexes: $\{c_1 c_3 c_{11}\},\{c_1 c_{10} c_{11}\},\{c_1 c_2 c_{10}\}, \{c_1 c_2 c_{11}\},\{c_2 c_5 c_{10}\},\{c_2 c_{10} c_{11}\} $. Next, the number of edges and triangles are normalized by the maximum number of edges and triangles, respectively. SCAE is then calculated as the negative logarithm of the ratio of the probability of having triangles and the probability of having edges. In this particular case, we have SCAE$=-\log[(6/165)/(10/55)]=-\log[0.2]=1.6$}
   \label{evol}
\end{figure}

\section*{Results}\label{Results and Discussion}
\subsection*{Simplicial complex entropy}
In dynamical systems, the estimation of the average rate of information creation is given by the Kolmogorov-Sinai (KS) entropy \cite{Kolmogorov1959, Sinai1959, LatoraAndBaranger1999}. However, in practice it is not applicable to real-world sequences because it requires a large amount of data. To overcome this limitation, several methodologies have been proposed (see \cite{Grassberger1983,Eckmann1985,Shaw1981} for an overview). In 1991, Pincus \cite{Pincus1991} introduced Approximate Entropy (ApEn) to assess irregularity in time series.
This measure depends on three parameters: the length of the time series $N$, the pattern length $m$ and the tolerance threshold $r$. ApEn measures the degree of irregularity or randomness of a time series: a lower value indicates more regularity, and a higher value represents more disorder or randomness.  Despite the usefulness of ApEn for studying various dynamic processes, modifications have been proposed to reduce potential biases, especially those related to counting the repetition of a pattern as coincident with itself.  Richman and Moorman \cite{RichmanAndMoorman2000} introduced the so-called sampling entropy (SampEn) to reduce the bias of ApEn. One of the advantages of SampEn is that it does not take into account self-matching and does not rely on a template-based approach (see \cite{Delgado_BonalAndMarshak2019,Humeau2015,FeutrillAndRoughan2021,JaminEtAl2020} for a recent survey of the topic). Following a similar approach to ApEn and SampEn, we introduce the statistics of entropy of a simplicial complex. We proceed as follows: given a time series $\{u_1,u_2,...,u_N$\}, we construct a $d$-dimensional space containing the vectors $c(i)=[u(i),u(i+1),...,u(i+d-1)]$ with $1\leq i \leq N-d+1$. From this point cloud we construct the Vietoris-Rips (V-R) complex (see Fig. \ref{evol} and Methods). We recall that the point cloud represents a $d$-dimensional space and the V-R complex are defined for a given radius $\epsilon$. Here, $\epsilon$ is used as  a factor of the standard deviation ($\sigma$) to define the effective radius, which allows comparing measurements of data sets with different amplitudes.
For simplicity, in the following we will focus on the case $d=2$. Let $s_k$ be the number of $k$-dimensional simplices present in the simplicial complex being analyzed.  Then we define $S_k= s_k/C({N-1},{k+1})$ as the probability that $k+1$ elements form a $k$-simplex, where $C(\cdot,\cdot)$ is the combination.  Similarly, we define $S_{k+1}=s_{k+1}/C({N-1},{k+2})$ as the probability that $k+2$ elements form a $(k+1)$-simplex. Next, the simplicial complex approximate entropy is defined as SCAE=$-\ln({S_{k+1}}/{S_{k}})$. This definition of SCAE has a direct interpretation, it is precisely the conditional probability that, for a given tolerance $\epsilon$, two $(k-1)$-simplex that form a $k$-simplex also  form part of a $(k+1)$-simplex. For instance, if $k=1$, we have the conditional probability that two points that are close each other and form an edge also form part of a triangle with a third point which is also close. Throughout this work, all the SCAE calculations have been performed using $k=1$, except when other cases are mentioned. 
As in the case of SampEn, a lower value of SCAE indicates a more regular behavior of the sequence whereas high values are assigned to more
irregular series. It is necessary to emphasize that SCAE is based on co-occurrences between pairs and triplets of patterns of the same length, which allows the analysis of configurations that are not considered in pairwise co-occurrences as in methods as ApEn or SampEn. 
We remark that SCAE depends on the length of  the time series $N$, the radius $\epsilon$ and the dimension $k$. Figure \ref{evol} illustrates how SCAE values are calculated.
\\
The SCAE can be extended to the case of two different time series in order to assess their degree of asynchrony/synchrony. In this case, given the time series $\{u_1,u_2,...,u_N$\} and $\{v_1,v_2,...,v_N$\}, the two-dimensional point cloud is constructed in terms of the vectors $c_{uv}(i)=(u(i),v(i))$ where $i$ ranges from $1$ to $N$. The  definition of  the C-SCAE is similar to the case of a single time series. Given a radius $\epsilon$, C-SCAE is given by -$\ln(S_{k+1}^{uv}/S_{k}^{uv})$, where $S_k^{uv}= s_k^{uv}/C({N},{k+1})$ and $S_k^{uv}= s_{k+1}^{uv}/C({N},{k+2})$ represent the probability that $k+1$ and $k+2$ elements form a $k$-simplex and ($k+1$)-simplex, respectively.   We observe that C-SCAE assigns a higher value when the joint asynchrony of the points is large (higher dispersion), while it is low when the points exhibit higher synchrony (low dispersion). \\

\subsection*{SCAE analysis of time series}
\begin{figure}[H]
\centering
\includegraphics[width=0.8\linewidth]{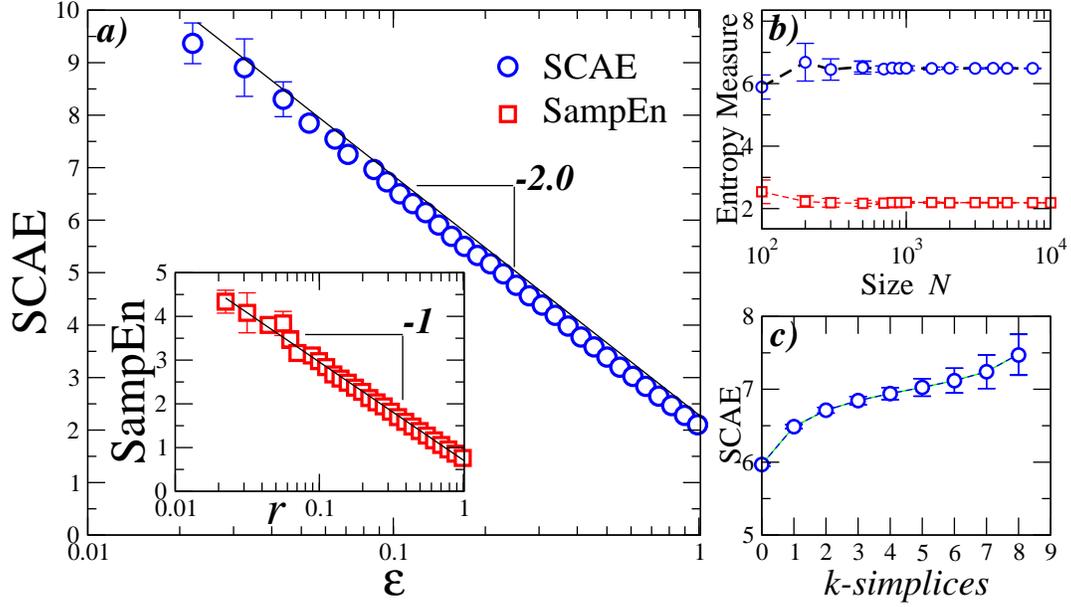}
\caption{Simplicial complex approximate entropy (SCAE) and Sample entropy (SampEn) of random numbers from a uniform distribution. a) SCAE as function of the radius $\epsilon$. SCAE values decay linearly with slope $\approx-2$ due to the fact that the horizontal axis is logarithmic. (Inset) SampEn as function of the radius $r$. SampEn also decays linearly in this semi-log plane but with slope $\approx-1$. 
We note that, according to the definition of V-R complexes, the $\epsilon$ radius would correspond to half the radius $r$ normally used in SampEn. 
The sequence lengths used for both entropies is $N=1024$. For SCAE we used $d=2$ and $k=1$, while for SampEn we used $m=2$. 
The straight lines represent the theoretical values given by SCAE$=2\ln(\epsilon_0/\epsilon)$, where $\epsilon_0=1/\sigma$ and SampEn$=\ln (r_0/r$), where $r_0=1/2\sigma$, with $\sigma$ the standard deviation of the original time series (see Methods). b) SCAE and SampEn as functions of system size $N$ for $\epsilon=0.1$ and $r=0.2$, respectively. c) SCAE for several dimensions of the simplicial complexes. In this case, we set $\epsilon=0.1$ and $N=2048$.  We observe that SCAE grows slowly as  $k$-dimension increases. Notice that when $k=1$, the SCAE values (for $\epsilon=0.1$) shown in panels a) and b)  are recovered.
The vertical bars at each value represent the standard deviation of 10 independent realizations.   }
\label{sampenvsscae}
\end{figure}

\textbf{Sequences from random numbers.} First, we apply SCAE on random numbers with a uniform distribution, and compare with the known results from SampEn. Fig. \ref{sampenvsscae}a shows the calculations of SCAE for iid random numbers as a function of $\epsilon$ with $N=1024$ points. Recall that $\epsilon$ is used as a factor of the standard deviation. We note that the SCAE values are almost three times higher than the SampEn values. For typical tolerance values such as $\epsilon=0.1$, the SCAE yields a value of $6.50 \pm 0.06$, while the SampEn leads to $2.20 \pm 0.03$. The range of possible values for SCAE can be useful when analyzing signals with noisy components and whose changes under different conditions are not easy to identify. SCAE decays linearly as a function of $\ln \epsilon$ with a negative slope close to -2. SampEn also decays linearly but with slope close to -1. The dependence of SCAE and SampEn on the length $N$ is depicted in Fig. \ref{sampenvsscae}b. Both measures exhibit independence of the system size as it increases.  
Next, we also examine the impact of the $k$-dimension on the SCAE. This dependence is shown in Fig. \ref{sampenvsscae}c, where it is observed that the SCAE grows slowly  as the dimension of the simplices increases.
Using the properties applicable to random numbers with uniform distribution, and the analytical approximation of the SampEn, it is possible to corroborate the numerical behavior of the SCAE (continuous line with slope -2) in the log-linear plane illustrated in Fig. \ref{sampenvsscae} (see Methods).\\

\begin{figure}[H]
\centering
\includegraphics[width=0.8\linewidth]{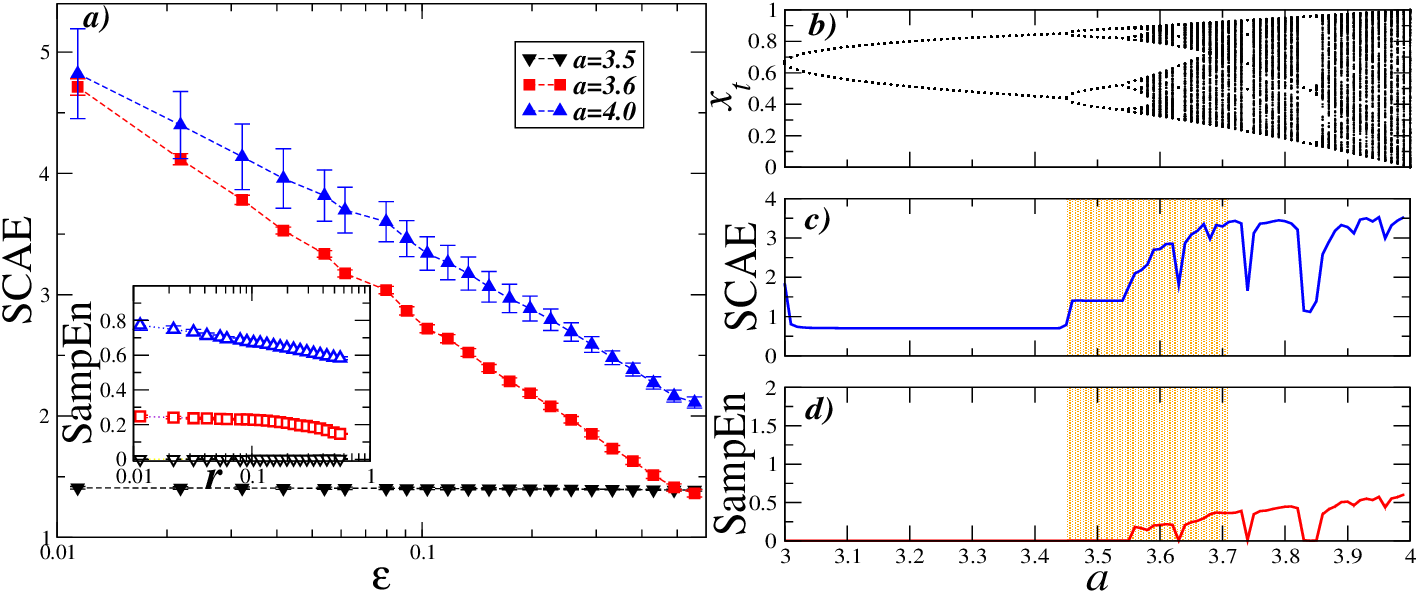}
\caption{Simplicial complex approximate entropy (SCAE) and  sample entropy (SampEn) for the logistic dynamics.
a) SCAE vs. radius $\epsilon$.  Here the SCAE calculations are shown with filled symbols, while SampEn values are indicated  by open symbols. For the periodic case ($a=3.5$), SCAE is constant and independent of $\epsilon$. As the parameter $a$ increases, SCAE exhibits a linear decay in the log-linear plot. Cleary, SCAE distinguishes between chaotic dynamics ($a=3.6$ and $a=4.0$) and marks a clear separation with periodic dynamics. For the calculations of SCAE, we used $k=1$, $d=2$ and $N=1024$. (Inset) SampEn vs. tolerance value $r$.  The values of SampEn corresponding to periodic dynamics ($a=3.5$) are zero; for chaotic dynamics they exhibit a slight variation as $r$ increases.  For the calculations of SampEn, we used $m=2$ and $N=1024$. The vertical bars at each value represent the standard deviation of 10 independent realizations. b) A bifurcation diagram of the logistic equation for several vaules of  the parameter $a$. (c) Average SCAE values obtained for $k=1$, $d=2$ and $\epsilon=0.1$. (d) Average SampEn values obtained for $m=2$ and $r=0.2$. The region shaded with light orange indicates the interval where SCAE is more sensitive to changes in $a$ compared to SampEn.      
}
\label{logistic}
\end{figure}

\begin{figure}[H]
\centering
\includegraphics[width=0.8\linewidth]{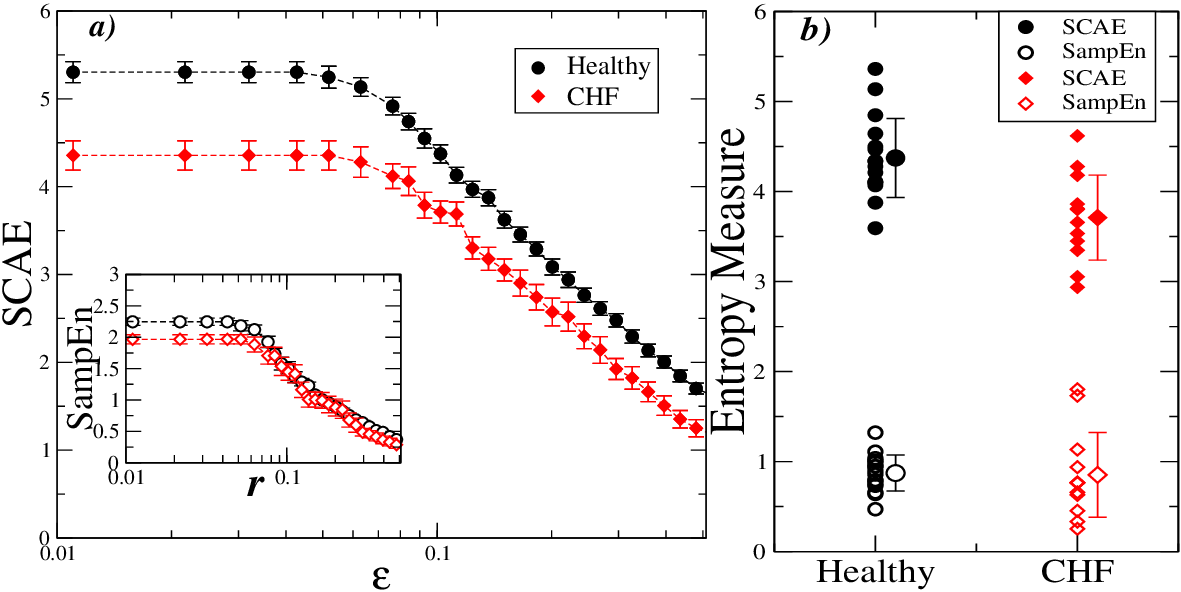}
\caption{Simplicial complex approximate entropy (SCAE) and  sample entropy (SampEn) for cardiac heartbeat time series.
a) SCAE vs. radius $\epsilon$.  SCAE analysis of RR time series from 16 healthy subjects and  12 patients with congestive heart failure (CHF). For each subject, $20$ segments with $1024$ non-overlapping data points were considered, then, the SCAE values from each segment were used to obtain an average value over a total of $20,480$ cardiac interbeat (RR) values. In these calculations, we used $k=1$ and $d=2$. The SCAE values from healthy subjects are significantly higher than from CHF patients. (Inset) SampEn vs. tolerance value $r$ for the same data in main frame. The vertical bars at each  value represent the standard error of the mean ($SEM=SD/\sqrt n$), where $n$ is the number of individuals. b) SCAE and SampEn for both groups using a specific tolerance value. We set $\epsilon=0.1$, $k=1$ and $r=0.2$, $m=2$ for SCAE and SampEn, respectively. Here, vertical bars at the mean value represent the standard deviation. For SCAE values, a significant difference is  observed between healthy and CHF groups ($p$-value $< 0.05$ by Student's test), while SampEn values of healthy subjects are not significantly different from CHF patients ($p$- value$=0.88$ by Student's test). }
\label{heartbeat}
\end{figure}

\begin{figure}[H]
\centering
\includegraphics[width=0.8\linewidth]{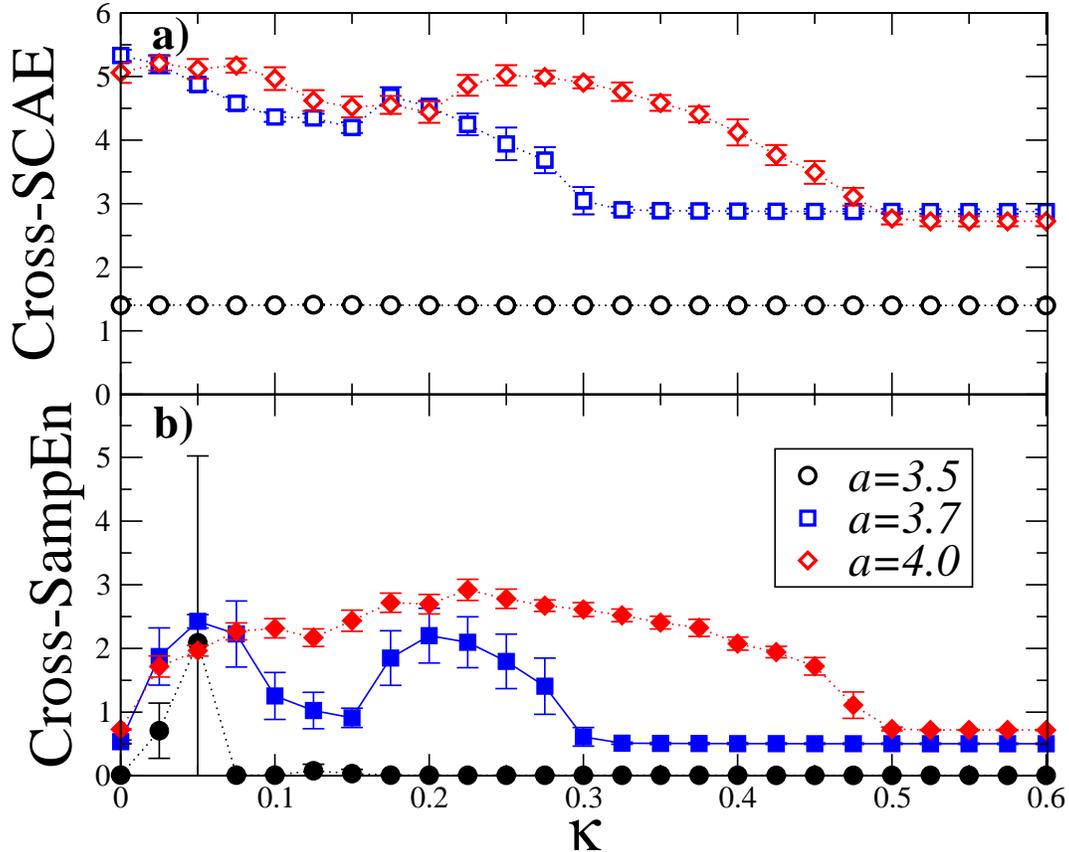}
\caption{Cross entropy measures vs. the coupling parameter $\kappa$ between two logistic maps. (a) Cross-SCAE  vs. $\kappa$ for periodic dynamics ($a=3.5$) and chaotic dynamics ($a=3.7$ and $a=4.0$). We used $k=1$, $\epsilon=0.1$ and $N=1024$.  
 b) As in a) but for SampEn vs. $\kappa$. Cross-SCAE has high values for low coupling ($\kappa<0.05$) while Cross-SampEn assigns low values, revealing that C-SCAE is genuinely capturing the asynchrony between the signals. As coupling increases, both measures capture transitions toward greater synchrony, except that C-SCAE shows greater differentiation between the cases. For the calculations of SampEn we set $m=1$, $r=0.2$ and $N=1024$. 
The vertical bars at each value represent the standard deviation of 10 independent realizations.   }
\label{crossscae}
\end{figure}

\textbf{Logistic chaotic dynamics.}
The SCAE definition is particularly useful for low dimensional nonlinear processes. To illustrate this, we have applied the SCAE to sequences obtained from the logistic map given by $x_{t+1}=ax_t(1-x_t)$. In particular, time series of $1024$ steps were generated for the values $a=3.5$, $3.6$, and $4.0$. $a=3.5$ corresponds to the periodic dynamics, while  $a=3.6 $  and $a=4.0$ produce chaotic dynamics. The results of the calculations of SCAE for different values of the radius $\epsilon$ are shown in  Fig. \ref{logistic}a. Clearly, SCAE distinguishes the different dynamics displayed by the system, specially those which correspond to the periodic ($a=3.5$) and the chaotic dynamics ($a=3.6$ and $4.0$). It is important to note that although this is a periodic dynamic ($a=3.5$), SCAE does not yield a value close to zero, as is the case with SampEn or ApEn. This provides an advantage for characterizing a wide variety of dynamics whose differentiation is potentially difficult with techniques such as SampEn. 
To further explore the transition from the quasi-periodic to the fully chaotic phase, both SCAE and SampEn were calculated for sequences generated with parameter $a$ within the interval $[3,4]$. Fig. \ref{logistic}b shows the bifurcation diagram where the route to chaos is illustrated by the asymptotic behavior of the variable $x_t$. The calculation results for SCAE and SampEn are presented in Fig. \ref{logistic}c and Fig. \ref{logistic}d, respectively. As the parameter $a$ changes, SCAE is more sensitive than SampEn, especially in the interval where the dynamics goes from a few stable cycles to a large diversity of them. 
\textbf{Heartbeat time series.}
As an example of applications of the SCAE to real-world time series, we apply our approach to cardiac heartbeat (RR) interval time series derived from electrocardiographic (ECG) recordings of healthy subjects and patients with congestive heart failure, a chronic condition associated with inefficient blood pumping  \cite{GoldbergerEtAl2000, BaimEtAl1986}. We analyze heartbeat interval time series from two groups: 16 healthy subjects (average age 32.6 years) and 12 patients with congestive heart failure CHF (average age 54.4 years) \cite{GoldbergerEtAl2000}. In our study, RR interval sequences with approximately $20\times 10^4$ beats corresponding to 4 diurnal hours of ECG records were considered. This database is part of an extended set of records which have been used in previous studies \cite{MietusEtAl2002,GoldsmithEtAl1997,IvanovEtAl1999,CostaPRE2005,Reyes-Ramirez2010}. 
For comparison and efficient data processing, for each subject we calculated the SCAE and SampEn for 20 non-overlapping segments of 1024 values and then considered the average of them.  In Fig. \ref{heartbeat}a, we present the results of the SCAE and SampEn calculations for several values of the radius.   Healthy subjects consistently exhibit a higher SCAE value with respect to the cardiac pathology (CHF) group for all $\epsilon$ values. In contrast, the results from the SampEn show very similar values between the two groups, making a differentiation difficult.  Fig. \ref{heartbeat}b shows a comparison between SCAE and SampEn for a specific value of the radius $\epsilon =0.1$. The SCAE values from the healthy group ( $4.37\pm0.43$, mean value $\pm$SD) are significantly ($t$-test with $p<0.05$) larger than those of CHF ($3.71\pm0.47$), while SampEn values (obtained with $r=0.2$) do not exhibit a significant difference ($t$-test with $p=0.88$).  
These SCAE-based results point in the direction that healthy dynamics are more complex than the fluctuations coming from dynamics in the presence of disease such as CHF, while SampEn is not able to identify significant differences between the same study groups. \\
\textbf{Logistic coupled maps.}
Next, we tested cross-SCAE to evaluate whether two series display a certain level of asynchrony. For this we consider two chaotic maps ( ($x_1(t)$ and $x_2(t)$) ), which are coupled according to the following relationship: $x_1(t+1)=(1-\kappa)f(x_1(t))+\kappa f(x_2(t))$ and $x_2(t+1)=f(x_2(t))$, with $f(x_i)=ax_i(1-x_i)$, $i=1,2$. Here $\kappa$ represents the coupling strength and $a$ is the parameter of the local logistic map. For $\kappa=0$, both maps evolve independently of each other, whereas for $\kappa=1$, $x_1$ is fully  coupled with $x_2$.  
We systematically evaluate the behavior of the Cross-SCAE for cases of periodic ($a=3.5$) and chaotic dynamics ($a=3.7$ and $a=4.0$) for several values of the coupling within the interval $[0,1]$. As shown in  Fig. \ref{crossscae}a, C-SCAE is relatively low for the periodic dynamics case ($a=3.5$) and is constant as $\kappa$ increases. 
Interestingly, C-SCAE reaches higher values for chaotic dynamics ($a=3.7$ and $a=4.0$) with zero coupling and then exhibits variations as $\kappa$ increases, to finally drop to moderate values and remain stable.  In particular, it is observed that C-SCAE, for the case $a=4.0$, reaches the state of highest synchrony when $\kappa=0.5$, while for the case $a=3.7$ this is reached when $\kappa=0.3$.
These results confirm that C-SCAE is capturing the changes in the coupling level even in the presence of chaotic dynamics. 
For comparison, Fig. \ref{crossscae}b shows the corresponding SampEn calculations for the same maps described above, and confirm that SampEn does not identify asynchrony for low coupling values between maps.



\section{Discussion}\label{Conclusions}
The method introduced in this work represents a reliable alternative to explore the complexity displayed by irregular time series. Our approach based on the occurrence of simplicial complexes has allowed us to evaluate the entropy for sequences of random values, with an acceptable correspondence between the analytical prediction and the numerical results.  Moreover, the proposed method is able to capture and quantify changes in the dynamics of chaotic maps, which represents an advantage over other standard methods for which quantification is more limited.  
When applied to the time series of intervals between heartbeats of healthy and congestive heart failure subjects, the SCAE results show that the healthy dynamics are more complex compared to those observed for the diseased group.  This result is in agreement with the idea that changes in cardiac dynamics under pathological conditions may be oriented towards either more regular fluctuations or greater irregularity.
We have extended our method to the bivariate case to study the level of synchrony between two time series, showing that entropy measures based on simplicial complexes, is able to capture details of  the transitions and levels of coupling between pairs of logistic maps.   \\
The results presented here agree with previous studies on the levels of irregularity in the signals, but have the particularity of not depending on larger dimensions, and have a wide range of possible values for a more detailed characterization.   Also, this approach naturally allows for multiscale analysis \cite{Costa2005}, which would provide an alternative way to evaluate changes across multiple scales in complex signals. 
\\
Our approach is focused on a two-dimensional point cloud, from which the VR complex is  constructed, but can be easily extended to higher dimensional spaces if necessary. This is natural for systems whose dimensionality is greater than 2, and where SCAE also allows a range of dimensions to be explored for simplicial complexes, potentially revealing this measure to be more informative of the underlying dynamics. Unlike the ApEn-based methods, SCAE retains the embedding dimension, which allows a clearer comparison between the systems.   
Also, although we have only shown SCAE results for complexes of order $k=1$ (and $k+1=2$), the statistics can be based on higher orders as long as the conditional probabilities are well defined. Further studies are needed to explore the feasibility of obtaining entropy measures at higher orders and whether multiscale procedures are applicable to both the univariate and bivariate cases \cite{FeutrillAndRoughan2021, JaminEtAl2020}.  
In addition, our method can be easily applied to sequences ranging from physiological to financial or geophysical records. \\ 
The main limitation of this study comes from the computational processing time for the SCAE calculation, where the construction of the simplicial complex is very demanding. Although these processes can be made more efficient by considering shorter data sets, they are still very time consuming. \\
We can conclude, therefore, that studying time series with a perspective of higher-order interactions (co-ocurrences) allows us to capture some important features, which are not revealed when analyzed by pairwise interactions (as occurs in the SampEn). In this sense, the presence or absence of higher-order co-ocurrences is indicative of the dynamics of the system, either with a certain level of regularity or completely random. 
Our method reveals additional details, which are not identified by other methods, and provides a way to obtain important information about complexity in stochastic and chaotic signals.  Furthermore, this study offers an alternative TDA-based methodology for the analysis of complex signals that can be easily extended to multiscale analysis to contribute to the understanding of the complexity of dynamical systems.

\section*{Methods}
\subsection*{Simplicial Complexes} 
The basic objects of Topological Data Analysis \cite{EdelsbrunnerAndHarer2008, EdelsbrunnerAndHarer2010, EdelsbrunnerAndMorozov2013} are the simplicial (abstract) complexes, which are mathematical structures used in algebraic topology to study the shape or structure of data.  We briefly define some properties of simplicial complexes and their filtered variants. 
Given a non empty set $V$, an abstract simplicial complex  $K$ is a collection of non empty subsets $\sigma\subseteq V$ subject to the following ``closed under subsets" condition: if $\sigma\in K$ and $\tau\subseteq\sigma$ then $\tau\in K$. An element $\tau\in K$ with $|\tau|=k+1$ is called a $k$-simplex. Thus, simplices with $1$ element are $0$-simplices, simplices with $2$ elements are $1$-simplices, etc. Formally, a $k$-simplex $\sigma$ is the covex hull of $k+1$ vertices affinely independent. Although this definition is rather abstract, it has a geometric interpretation (which, in fact, can be made in a formal connection, which for brevity we omit here): $0$-simplices correspond to points in euclidean space, $1$-simplices correspond to edges, $2$-simplices correspond to triangles, and generally $k$-simplices correspond to convex hulls of collections of $k+1$ points in so-called general position in euclidean space; the ``closed under subsets" condition similarly can be interpreted geometrically; for instance, faces of 2-simplices, that is, triangles, are edges, which in turn are $1$-simplices. 
A \emph{filtered} simplicial complex $K$ is a collection of simplicial complexes $\{K_i\}_{i\in I}$, where $I$ is a totally ordered set such that for $i,j\in I$ with $i\leq j$ we have $K_i \subseteq K_j$. 
In practice, however, we often have that $I$ are nonnegative integers and $K_i \subsetneq K_{i+1}$ only occurs for a finite number of indices in $I$. Intuitively, we are adding a finite number of simplices to an initial complex over time until we arrive at a final complex. A particular type of filtered simplicial complex corresponds to the so-called Vietoris-Rips (VR) complex, $K$, which is associated to a point cloud $C=\{c_1,\ldots,c_N\}$. Briefly, the VR complex is defined as follows:
Given $\epsilon>0$, define $K_{\epsilon}$ by considering all simplices $\sigma = \{x_1, \ldots, x_l\}\subseteq C$ such that $\lVert x_i - x_j \rVert_2<\epsilon$ for all $x_i, x_j$ in $\sigma$. A straightforward verification shows that $K_{\epsilon} < K_{\epsilon^{'}}$ if $\epsilon<\epsilon^{'}$. VR complexes have theoretical properties that have been well documented \cite{Zomorodian2012}. In particular, they model closely the shape of data point clouds in euclidean space (more specifically, they approximate the so-called Filtered C\v{e}ch complex associated to a point cloud, which in turn models the homotopy type of the union of balls centered at the points $c_i$). Detailed theoretical information about homology and related concepts can be consulted in  \cite{Munkres1984, SeifertAndThrelfall1980, Hatcher2001, EdelsbrunnerEtAl2002, Zomorodian2010, Zomorodian2012, Munkres2014, DeyAndWang2022}, among others. 
In this paper, we restrict ourselves to construct the VR complexes in a two-dimensional space for a predefined $\epsilon$ radius (Fig. \ref{evol}b).  Once the point cloud has been constructed, the VR complexes are determined using the \texttt{Python} package \texttt{Gudhi} \cite{Gudhi}. From these complexes, we perform the quantifications of the presence of the complexes of order $k$ to determine the conditional probabilities considered in the definition of the SCAE. 

\subsection*{Theoretical analysis of the SCAE for random numbers.}We describe the analytical derivations of SCAE for the case of random numbers with uniform distribution. 
First, we considered the case where $d=1$, i.e., we have a one-dimensional point cloud. For the case $k=0$ ($0$-simplex), the SCAE is  defined as  the negative logarithm of the conditional probability that the distance between two points (i.e. two $0$-simplex) is less or equal to  $\epsilon$ given that we have a single point. Since there is no correlation between the points, it is possible to show that the SCAE simply equals the negative logarithm of the probability that two points are at a distance less than or equal to $\epsilon$. More formally,  for random numbers with density function $p(x)$ and standard deviation $\sigma$, we have SCAE$(k=0,d=1)=-\ln [\int p(x)(\int_{x-\epsilon\sigma}^{x+\epsilon\sigma}p(z)dz)dx]\sim \ln(\epsilon_1/\epsilon)$, with $\epsilon_1=1/2\sigma$.  It is important to note that this case corresponds to that of the SampEn definition for random values (see details in Refs. \cite{Pincus1991, RichmanAndMoorman2000, Costa2005}).
Next, we analyze the case $d=2$. For $k=1$, SCAE$(k=1,d=2)$ is given by the negative logarithm of  the conditional probability that three points are close (within $\epsilon$) each other given that the distance between two of the three points is less than or equal to $\epsilon$. By applying the fact that the points are uncorrelated, it is possible to show that  SCAE$(k=1,d=2)=-\ln [\int p(x)(\int_{x-(\epsilon/2)\sigma}^{x+(\epsilon/2)\sigma}p(z)\{\int_{z-(\epsilon/2)\sigma}^{z+(\epsilon/2)\sigma}p(y)dy\}dz)dx]= -\ln(\sigma\epsilon)^2=2\ln(\epsilon_0/\epsilon)$, with $\epsilon_0=1/\sigma$. 
These results can be generalized for higher dimensions $d\geq 2$, to obtain SCAE$(d)=-\ln [\int p(x)\{\int_{x_1-(\epsilon/2)\sigma}^{x_1+(\epsilon/2)\sigma}p(z_1)dz_1\cdot\cdot\cdot \int_{x_j-(\epsilon/2)\sigma}^{x_j+(\epsilon/2)\sigma} p(z_j)dz_j\}dx]$, where the sub-index $j$ runs from $j=1$ to $j=d$. Then, the SCAE$(d)= -\ln(\sigma\epsilon)^d=d\ln(\epsilon_0/\epsilon)$,  with $\epsilon_0=1/{\sigma}$.
The values obtained (for $d=2$) with the expression SCAE$(d=2)=2\ln(\epsilon_0/\epsilon)$  are plotted in Fig. \ref{sampenvsscae}.   

\section*{Data availability}
The experimental data used in this study are publicly available online at \hyperref[https://www.physionet.org]{www.physionet.org} 

\bibliography{references}

\section*{Acknowledgements}
 We thank P. Lara, B. Obregon, C. Masoller and L. Liebovitch for useful discussions and suggestions. This  work  was  partially  supported  by  programs  EDI  and  COFAA  from  Instituto Polit\'ecnico Nacional and Consejo Nacional de Ciencia y Tecnolog\'ia, M\'exico. 

\section*{Author contributions statement}
L. G.V. conceived and designed the experiments; L.G.V., A.Z.O. and A.G.S.  analysed the results.  L.G.V. and A.G.S. wrote the paper. All authors reviewed the manuscript.  
\section*{Corresponding author}
Correspondence to Lev Guzmán-Vargas (lguzmanv@ipn.mx)
\section*{Competing Interests}

The authors declare no competing interests.



\end{document}